\documentclass[5p,times]{elsarticle}

\usepackage{ecrc}

\volume{00}

\firstpage{1}

\journalname{Computer Methods and Programs in Biomedicine}

\runauth{}

\jid{CMPB}

\jnltitlelogo{Computer Methods and Programs in Biomedicine}

\usepackage{amssymb}
\usepackage[figuresright]{rotating}
\usepackage[utf8]{inputenc}
\usepackage{xcolor}
\usepackage{natbib}
\usepackage{graphicx}
\usepackage{multirow}
\usepackage{amsmath}
\usepackage{amssymb}
\usepackage{amsfonts}
\usepackage{environ}
\usepackage{xfrac}
\usepackage{caption} 
\begin{document}

\begin{frontmatter}

\dochead{}

\title{An Hybrid Method for the Estimation of the Breast Mechanical Parameters}

\author[l1]{Diogo Lopes}
\author[l2]{Stéphane Clain}
\author[l1]{António Ramires Fernandes}

\address[l1]{Centro~Algoritmi,~Universidade~do~Minho}
\address[l2]{Centro de Física,~Universidade~do~Minho}

\begin{abstract}

\subsection*{Background}
There are several numerical models that describe real phenomena being used to solve complex problems. For example, an accurate numerical breast model can provide assistance to surgeons with visual information of the breast as a result of a surgery simulation. 
%The process of finding the model parameters for a particular breast requires inputs, either in the form of medical imaging techniques or distance and volume measures. 
The process of finding the model parameters for a particular breast requires numeric inputs, either based in medical imaging techniques, or distance and volume measures. 
Inputs can be processed by iterative methods (inverse elasticity solvers). Such solvers are highly robust and provide solutions within the required degree of accuracy. However, their computational complexity is costly in terms of time. On the other hand, machine learning based approaches, such as Multilayer Neural Networks (MNN), provide outputs in real-time. Although high accuracy rates can be achieved, these methods are not exempt from producing solutions outside the required degree of accuracy. In the context of real life situations, a non accurate solution might present complications to the patient as well as to the surgeon.

\subsection*{Methods}
We present an hybrid parameter estimation method to take advantage of the positive features of each of the aforementioned approaches. Our method preserves both the real-time performance of deep-learning methods, and the reliability of inverse elasticity solvers. The underlying reasoning behind our proposal is the fact that deep-learning methods, such as neural networks, can provide accurate results in the majority of cases and they just need a fail-safe system to ensure its reliability. Hence, we propose using a MNN to get an estimation which is in turn validated by a iterative solver. In case the MNN provides an  estimation not within the required accuracy range, the solver refines the estimation until the required accuracy is achieved.

\subsection*{Results}
Our tests show that in the vast majority of cases the MNN is capable of achieving an accurate solution on its own. When such a solution is outside the required accuracy this solution will act as the initial estimate for the iterative solver (fail-safe system) to further refine it. In these latter case, the solution tends to be close enough for the iterative solver to be able to refine in a short time span.

\subsection*{Conclusions}
%Based on the results we obtained we can conclude that the hybrid method is both fast and reliable. 

Based on our results we can conclude that the hybrid method is able to complement the computational performance of MNNs with the robustness of iterative solver approaches, providing both a fast and accurate method that can be applied to a vast range of problems.

\end{abstract}

\begin{keyword}

Hybrid Numerical Estimation, Multilayer Neural Network, Numerical Model, Breast Parameter Estimation, , Machine Learning, Iterative Solver

\end{keyword}

\end{frontmatter}

% \linenumbers

\section{Introduction}
\label{deep::intro}

The usage of iterative solvers to solve numerical problems, such as inverse elasticity, is very common in literature (\cite{Delingette2004}). Most of them are the result of combining various methods and techniques that in the end provide a solution (\cite{Delingette2004}). The application of these solvers in real problems showed their usefulness and they became regarded as reliable methods to solve said problems (\cite{Delingette2004, tromeur2006, Palomar2008, Rajagopal2008, mohamed2018}).

Despite their reliability, these methods are considered costly in terms of time consumption (\cite{Azar2002, Delingette2004, tromeur2006, Lopes2018}). The time consumption is directly related to the number of iterations required to find a solution to the problem (\cite{Delingette2004, tromeur2006}), the number of iterations being highly dependent on the initial estimation
%. In fact, many of these methods like the Gauss-Newton, depend on an initial estimation of the final result which if it is very close, than the method will only require a few iterations 
(\cite{Delingette2004, marques2016, Lopes2018}). With that in mind, several approaches were made to develop methods capable of providing these solvers with better approximations as initial estimations which could accelerate the convergence and therefore reduce the time spent with the solver (\cite{tromeur2006, barabasz2014, simoncic2015, marques2016, Lopes2018}).

These types of problems can be seen as regression problems, and on that account, machine learning methods such as MNNs, started to be an alternative and viable approach in several fields from computer science to biology (\cite{Hinton2006,LeCun2015,DePristo2016,Goodfellow2016, Zhou2018}). In fact, there is already available literature where they are used to help solve real life problems involving iterative methods (\cite{Litjens2017, Sun2017, Martinez2017, Hamidinekoo2018, rezaee2019}). 

The main issue with MNNs and other machine learning methods is that they are not 100\% reliable and can occasionally produce outputs outside the required safety margin. If we consider problems where the health of a patient is involved, than, the usage of these methods becomes very limited, since using inaccurate data can potentially lead to health complications.

Lets consider as a case study, the iterative solver proposed by \cite{Lopes2018} that is used to estimate the biomechanical parameters of the breast. The method receives a set of breast measurements and then starts an iterative process from an initial estimation of the breast parameters that concludes when the measurements produced by the estimated parameters match (to a certain degree) the input breast measurements (\cite{Lopes2018}). Despite the work proposing ways to accelerate the convergence, the method can, for some cases, need hours to provide an accurate solution (\cite{Lopes2018}).

We propose a new hybrid method to solve estimation/regression problems, that combines the capacity of obtaining nearly instantaneous results by a machine learning method (MNN) and the robustness and accuracy of iterative solvers. This method uses data from the known problem to train a neural network which will then be used to provide a solution to the problem. Then, the iterative solver will validate the provide solution. If the solution does not meet the accuracy requirements, than the iterative solver refines the aforementioned solution.

Note that, the intent behind this hybrid method is not to use the MNN as a rough initial estimator as what happens in other studies (\cite{Martinez2017, rezaee2019}), but to use the MNN as a replacement for the slow iterative solvers understanding at the same time its limitations.

We use as case study the iterative solver mentioned above (\cite{Lopes2018}) and will conduct several tests. The results obtained show that this method achieved the goals of becoming faster (MNN working $\approx 99\%$ of times) and robust (the iterative solver was able to refine the breast parameters when the MNN solution was not accurate enough).

Subsection \ref{breast::model} (in \ref{sec:methods}) details the iterative method (case study) used to demonstrate the validity of the hybrid approach. Section \ref{sec:results} presents and evaluates the neural network approach comparing it with an iterative method (subsection \ref{sec:nn}) and the hybrid approach is described in subsection \ref{sec:hybrid}. Finally, in \ref{sec:conclusion} we discuss the general results and present some conclusions and possible avenues for future work.

\section{Methods}
\label{sec:methods}

\subsection{Case Study: Breast Model}
\label{breast::model}

We provide a brief description of the iterative method used to demonstrate the advantages of the hybrid approach. The method presented in \cite{Lopes2018} was selected because it is an analytical method and a large set of examples can be generated both for training, and testing the neural network.

The breast is a complex structure constituted of a mass of glandular tissue encased in fat that accounts for its characteristic round shape, being connected to the skin through a series of ligaments. These tissues possess different bio-mechanical properties depending of the patient (age, fat layer length, skin elasticity among others), and as such behave differently to external perturbations, namely the gravity. Deformation evaluation of the breast over these external actions is achieved by considering a simplified stress-free geometrical domain of the breast equipped with a Neo-hookean mechanical model (\cite{Palomar2008}) where the breast inner tissues and the skin are discretised (as in \cite{Lopes2017,Lopes2018}).

\begin{figure}[ht]
\includegraphics[width=0.25\textwidth]{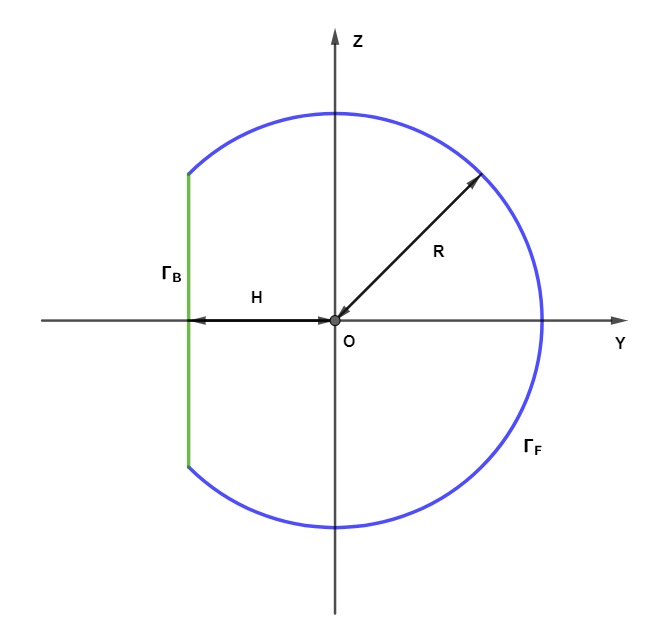}
\includegraphics[width=0.21\textwidth]{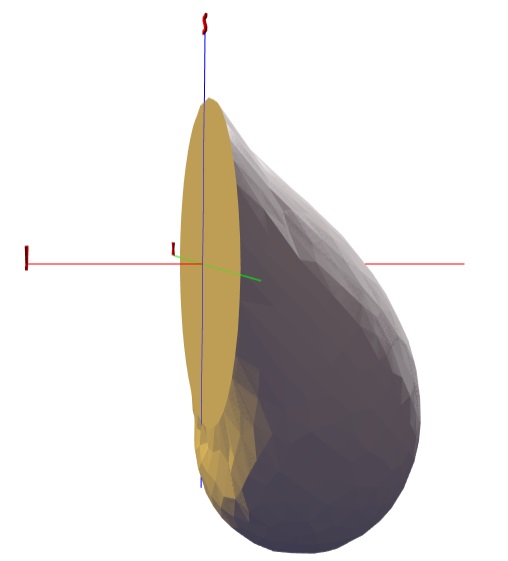}
\centering
\caption{Breast scheme with the geometrical parameters (left) and digital breast example under gravity (right) - \cite{Lopes2018}}
\label{fig:breast_geo_mech}
\end{figure}

The breast's shape and size in a stress-free domain consists in a spherical cap where the plane section is attached to the torso. This geometrical structure is defined by two parameters: the radius $R$ and the truncated length of the cap $H$. The breast visco-elastic properties are defined by a set of four mechanical parameters: $\lambda_{br}$ and $\mu_{br}$ for the glandular and fat tissues, and $\lambda_{sk}$ and $\mu_{sk}$ for the breast skin. The geometrical parameters are denoted by $\Lambda_{g}=(R,H)$ while the mechanical parameters are denoted by $\Lambda_{m}=(\lambda_{br},\mu_{br},\lambda_{sk},\mu_{sk})$ and they can be used to generate a digital breast (figure \ref{fig:breast_geo_mech}).

\begin{figure}[ht]
\includegraphics[width=0.35\textwidth]{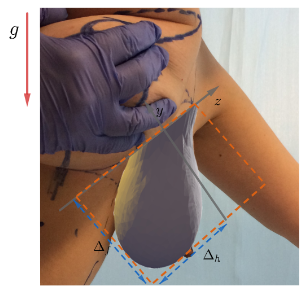}
\centering
\caption{Scheme of breast measurements of a patient - \cite{Lopes2018}}
\label{fig:real_measuring}
\end{figure}

The six parameters $\Lambda=(\Lambda_{g},\Lambda_{m})$ that characterise the breast shape, size, and the mechanical characteristics have to be determined for each patient. In \cite{Lopes2018}, they use 15 measurements $M=(M_1,\cdots,M_{15})$ to estimate these six parameters. The aforementioned measurements consist on the breast's volume, skin surface area, breast height, frontal and back depth for the patient in three different positions depicted in Figure \ref{fig:real_measuring} (see \cite{Lopes2018} for the details).

We aim at finding the set of parameters $\Lambda$ from a given set of measurements $M$ carried out on a specific patient. There are several methods proposed in literature that provide the parameters' identification operator $M \to \Lambda(M)$ (\cite{Azar2002,Palomar2008,Cardoso2010,Lopes2018}). We choose the method presented in \cite{Lopes2018} due to the ease of reproducing its values (does not require medical imaging data and it allows to simulate several different breast configurations).

The results obtained with this method showed a good accuracy in terms of the estimation of the parameters. The main issue is the large amount of time (in some cases it can take hours \cite{Lopes2018}) necessary to estimate the breast parameters.

\subsection{MNN Estimator}
\label{sec:nn}

We present a MNN model and detail the training procedure we carry out to emulate the inverse problem solver, \textit{i.e.} to provide an approximation of $\Lambda$ as a function of measurements $M$. Similarly to \cite{Lopes2018}, the main goal of the breast parameter estimation method is the determination of a set of parameters that produces a relevant digital model of the real breast, \textit{i.e.} a model which accurately reproduces the real breast mechanical behaviour as well as its aesthetics. Such model is then used to support a surgeon decision making regarding surgical procedures and, hence, reduces the risk of errors.

Although accurate, inverse solver methods are too time consuming to be of practical use. Despite several code optimisations, the running times reported in \cite{Lopes2018} can go up to several hours of computation whereas MNNs can provide an answer in real-time.

\begin{figure}[ht]
\includegraphics[width=0.45\textwidth]{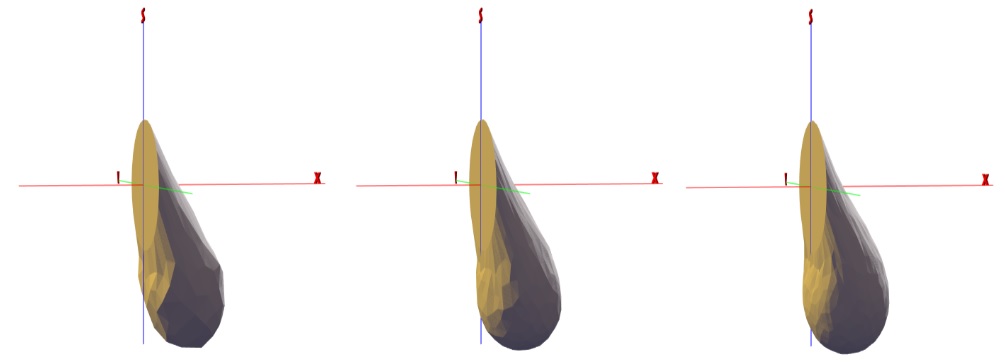}
\centering
\caption{Image of 3 breast meshes generated by the same parameters: coarse (left), medium (mid) and thin (right)}
\label{fig:breast_granularity}
\end{figure}

To evaluate the accuracy of the proposed approach we will compare our results with the values presented in \cite{Lopes2018} using three different meshes: coarse, medium and thin. The visual difference between these mesh resolutions is depicted in figure \ref{fig:breast_granularity}. 

To train, validate and test the MNNs, datasets were created using the numerical breast model proposed in \cite{Lopes2017} carrying out numerical simulations with a wide range of variation of coefficients $\Lambda$. We elaborate a dataset for each mesh (${\mathcal D}_{coarse}$, ${\mathcal D}_{medium}$ and ${\mathcal D}_{thin}$) with $N=50000$. Each dataset ${\mathcal D}_{k},\ k \in \{coarse,medium,thin\}$ is split into a test set with 10000 cases (${\mathcal D}^{test}_{k}$), a validation set with 4000 samples (${\mathcal D}^{val}_{k}$), the remaining 36000 cases being the training set (${\mathcal D}^{train}_{k}$).

\subsubsection{Datasets generation}\label{nn:data_generation}
To generate the datasets, random samples are drawn from a Gaussian distribution of the parameters centred on a set of values suggested by surgeons as representing an average looking breast $\Lambda^{average}=(R=0.05m$, $H=0.0564m$, $\lambda_{br}=1000Pa$, $\lambda_{sk}=8000Pa$, $\mu_{br}=400Pa$ and $\mu_{sk}=1600Pa)$. 
Then a dataset is created as follows:
\begin{enumerate}
    \item Generate $N$ valid vectors $s^n\in \mathbb R^6$, $n=1,\cdots,N$ such that each component $s^n_i$ is chosen randomly with the Normal law $\mathcal{N}(0,\,1/2)$. Set $\Lambda^n_i=(1+s^n_i)\Lambda^{average}_i$, $i=1,\cdots,6$, $n=1,\cdots,N$. A vector is considered invalid when some of the parameters, or a combination of them,  do not represent plausible breast models \cite{Lopes2017};
    
    \item  Compute the MNN inputs $M^n=(M^n_1,\cdots,M^n_{15})$ corresponding to the MNN outputs $\Lambda^n=(\Lambda_1^n,\cdots,\Lambda^n_6)$, using the method detailed in \cite{Lopes2018} and create the elements $D^n = \{M^n,\Lambda^n\}$. %($M^n$)
    
    \item When the dataset is fully generated, its elements $D^n$ are normalised as follows:
    \begin{align*}
        M^n_j &:=\frac{M^n_j}{\overline M_j},\ j=1,\cdots,15,\qquad \\
        \Lambda^n_i &:=\frac{\Lambda^n_i}{\overline \Lambda_i},\ i=1,\cdots,6,    
    \end{align*}
    where $\overline M_j$ and $\overline \Lambda_i$ stand for the mean values over the whole dataset.

\end{enumerate}

%%%%%%%%%%%%%%%%%%%%%%%%%%%%%%%%%%%%%%%%%%%%%%%%%%%%%%%%%
%%%%%%%%%%%%%%%%%%%%%%%%%%%%%%%%%%%%%%%%%%%%%%%%%%%%%%%%%
\subsubsection{MNN Configuration}
\label{nn:analysis}

A MNN configuration relates to its architecture, the activation functions and on the optimiser, among other parameters. We experimented several configurations for the MNN and found that the best results were obtained with a MNN constituted of 12 hidden layers with 128 nodes per layer, using the exponential linear unit activation function for each layer, and the RMSProp optimiser. The input layer has then fifteen inputs for the measurements to provide an output layer of six nodes for the geometric and mechanical parameters.

%%%%%%%%%%%%%%%%%%%%%%%%%%%%%%%%%%%%%%%%%%%%%%%%%%%%%%%%%
%%%%%%%%%%%%%%%%%%%%%%%%%%%%%%%%%%%%%%%%%%%%%%%%%%%%%%%%%
\subsubsection{Learning}
\label{nn:analysis:learn}
We trained a MNN for each different sized meshes and tested their accuracy. Implementation was carried out in Python using Google's Tensorflow with Keras (\cite{tensorflow2015, chollet2015keras}). The values per mesh reported in this section are an average of 5 runs. Each run is considered complete after it ran 3000 epochs and the best epoch regarding the validation set is selected. It is important to mention that the training and validation datasets are a result of splitting of a dataset of 40000 cases at each run where $90\%$ (36000 cases) goes for training and the remaining $10\%$ (4000 cases) goes for validation.

We draw comparisons between the original Inverse Method (IM) and the Neural Network approach. In table \ref{table::nn:train_eval_res}, we report on the MNN performance for both the training and validation sets, together with the results in \cite{Lopes2018} for the IM approach (relative error with respect to the exact value given in \%).
\begin{table}[ht]
\setlength{\tabcolsep}{1.5pt}
\renewcommand{\arraystretch}{1.5}
\centering
\small
\begin{tabular}{|c|c|c|c|c|c|c|c|c|c|}
\hline
\multicolumn{3}{|c|}{\textbf{\begin{tabular}[c]{@{}c@{}}Error per \\ Parameter\\ (\%)\end{tabular}}} & \textit{\textbf{$R$}} & \textit{\textbf{$H$}} & \textit{\textbf{$\lambda_{br}$}} & \textit{\textbf{$\lambda_{sk}$}} & \textit{\textbf{$\mu_{br}$}} & \textit{\textbf{$\mu_{sk}$}} & \textit{\textbf{\begin{tabular}[c]{@{}c@{}}avg\\ total\end{tabular}}} \\ \hline
\multirow{3}{*}{\textbf{Coarse}} & \multicolumn{2}{c|}{\textit{\textbf{IM}}} & 2.30 & 2.71 & 5.28 & 9.41 & 3.07 & 15.72 & 6.42 \\ \cline{2-10} 
 & \multirow{2}{*}{\textit{\textbf{MNN}}} & \textit{\textbf{train}} & 0.73 & 1.08 & 2.19 & 3.53 & 1.28 & 11.02 & 3.31 \\ \cline{3-10} 
 &  & \textit{\textbf{val}} & 0.76 & 1.09 & 2.31 & 3.81 & 1.35 & 11.70 & 3.50 \\ \hline
\multirow{3}{*}{\textbf{Medium}} & \multicolumn{2}{c|}{\textit{\textbf{IM}}} & 2.42 & 3.78 & 5.11 & 9.24 & 2.68 & 15.19 & 6.40 \\ \cline{2-10} 
 & \multirow{2}{*}{\textit{\textbf{MNN}}} & \textit{\textbf{train}} & 0.82 & 1.35 & 2.15 & 2.98 & 0.98 & 8.38 & 2.78 \\ \cline{3-10} 
 &  & \textit{\textbf{val}} & 0.86 & 1.40 & 2.20 & 3.04 & 0.99 & 8.61 & 2.85 \\ \hline
\multirow{3}{*}{\textbf{Thin}} & \multicolumn{2}{c|}{\textit{\textbf{IM}}} & 2.56 & 3.83 & 5.07 & 9.13 & 2.48 & 14.43 & 6.25 \\ \cline{2-10} 
 & \multirow{2}{*}{\textit{\textbf{MNN}}} & \textit{\textbf{train}} & 1.24 & 3.00 & 3.97 & 3.80 & 1.18 & 9.40 & 3.77 \\ \cline{3-10} 
 &  & \textit{\textbf{val}} & 1.23 & 3.11 & 3.94 & 4.11 & 1.22 & 9.54 & 3.86 \\ \hline
\end{tabular}
\caption{Error values in percentage for the iterative method (IM) and the multilayer neural network (MNN). The MNN values are presented for training (train) and validation (val). The error values are presented per parameter and total average for the three mesh sizes. }
\label{table::nn:train_eval_res}
\end{table}

The training results show that the MNNs are capable of associating the set of given measurements $M$ to the breast model parameters $\Lambda$ with different degrees of success depending on the parameters and on the mesh size. Comparing the errors per parameter of the MNN and the IM, we observe that the MNN approach obtains a significantly better approximation for every parameter. Such an improvement comes from the averaging character of the training that reduces the random component of the error contained in the data. In particular, the MNN provides the best approximations for the geometrical parameters ($R$ and $H$) with only $0.73\%$ and $1.08\%$ error respectively with a coarse mesh and $1.24\%$ and $3.00\%$ error respectively with a thin mesh. Regarding the mechanical parameters ($\lambda_{br}$, $\mu_{br}, \lambda_{sk}$ and $\mu_{sk}$), the MNN provides the best approximations with the medium mesh size. 

A closer analyse of the results for each parameter shows that the geometrical parameters $R$ and $H$ as well as the mechanical parameters of the breast bulk tissue $\lambda_{br}$ and $\mu_{br}$ are well-estimated with very small errors (values around $2\%$ or lower). The mechanical parameters of the skin present larger errors with approximately $3\%-4\%$ error for $\lambda_{sk}$ and approximately $8\%-11\%$ error for $\mu_{sk}$. The same difficulties in estimating $\mu_{sk}$ were observed in \cite{Lopes2018}. In fact, they pointed out that, even at the numerical level, \textit{i.e.}, using the iterative inverse solver, it is difficult to obtain a good approximation value (relative error larger than $14\%$ for $\mu_{sk}$) due to a very sensitivity of this parameter to the measure errors.  

Taking into consideration the values obtained with the IM, it could be expected that the thinner the mesh, the better the approximation obtained with the MNN. However, \cite{Lopes2018} stated that the relation between the parameters and the measurements is nonlinear.
This means that any error estimating a certain parameter will affect the other estimated parameters disproportionately (\cite{Lopes2018}). Therefore, using very coarse or very thin meshes can compromise the overall estimation of the breast parameters, and a possible optimal mesh size can be found where it balances the geometrical aspects of the breast and the detail for the mechanical parameters (\cite{Lopes2018}).

The validation values of table \ref{table::nn:train_eval_res} are similar to the training values.% (even if slightly worse as expected).
They present values of average error per parameter of $3.5\%$ for a coarse mesh, $2.85\%$ for a medium mesh and $3.86\%$ for a thin mesh. These results are very good when compared with the ones obtained with the IM whose average error per parameter was $6.42\%$ for a coarse mesh, $6.4\%$ for a medium mesh and $6.25\%$ for a thin mesh.

\section{Results}
\label{sec:results}

\subsection{Multilayer Neural Networks}

%%%%%%%%%%%%%%%%%%%%%%%%%%%%%%%%%%%%%%%%%%%%%%%%%%%%%%%%%
%%%%%%%%%%%%%%%%%%%%%%%%%%%%%%%%%%%%%%%%%%%%%%%%%%%%%%%%%
\subsubsection{MNN Performance}
\label{nn:analysis:test}

We assess the efficiency of the training MNN model using the test datasets with 10000 cases for each mesh resolution. The results for the MNN model and the IM method are presented in table \ref{tab:nn:it:rob_50} showing the relative errors as well as the time spent by each method. The values for the IM are the same as the ones presented in table \ref{table::nn:train_eval_res} in section \ref{nn:analysis:learn}. 

\begin{table}[ht]
\setlength{\tabcolsep}{1.0pt}
\renewcommand{\arraystretch}{1.25}
\centering
\begin{tabular}{|c|c|c|c|c|c|c|c|c|c|c|}
\hline
\multicolumn{2}{|c|}{\textbf{\begin{tabular}[c]{@{}c@{}}Error per \\ Parameter\\ (\%)\end{tabular}}} & \textit{\textbf{$R$}} & \textit{\textbf{$H$}} & \textit{\textbf{$\lambda_{br}$}} & \textit{\textbf{$\lambda_{sk}$}} & \textit{\textbf{$\mu_{br}$}} & \textit{\textbf{$\mu_{sk}$}} & \textit{\textbf{\begin{tabular}[c]{@{}c@{}}avg\\ total\end{tabular}}} & \textit{\textbf{cycles}} & \textit{\textbf{\begin{tabular}[c]{@{}c@{}}time\\ (min)\end{tabular}}} \\ \hline
\multirow{2}{*}{\textbf{Coarse}} & \textit{\textbf{IM}} & 2.30 & 2.71 & 5.28 & 9.41 & 3.07 & 15.72 & 6.42 & 12 & 3.27 \\ \cline{2-11} 
 & \textit{\textbf{MNN}} & 0.77 & 1.10 & 2.34 & 3.85 & 1.39 & 11.77 & 3.54 & - & $3e^{-4}$ \\ \hline
\multirow{2}{*}{\textbf{Medium}} & \textit{\textbf{IM}} & 2.42 & 3.78 & 5.11 & 9.24 & 2.68 & 15.19 & 6.40 & 13 & 11.6 \\ \cline{2-11} 
 & \textit{\textbf{MNN}} & 0.88 & 1.40 & 2.24 & 3.09 & 1.02 & 8.70 & 2.89 & - & $3e^{-4}$ \\ \hline
\multirow{2}{*}{\textbf{Thin}} & \textit{\textbf{IM}} & 2.56 & 3.83 & 5.07 & 9.13 & 2.48 & 14.43 & 6.25 & 14 & 26.45 \\ \cline{2-11} 
 & \textit{\textbf{MNN}} & 1.25 & 3.07 & 4.01 & 4.13 & 1.27 & 9.56 & 3.88 & - & $3e^{-4}$ \\ \hline
\end{tabular}
\caption{Comparison of the two parameter estimations methods: Multilayer Neural Network (MNN) and the Iterative Method (IM). Mean difference in percentage between the reference parameters and the estimated parameters using three mesh sizes: Coarse, Medium and Thin. Time spent by each method in average and the number of iterations (cycles) required by the iterative method to converge and produce a good result.}
\label{tab:nn:it:rob_50}
\end{table}

Test errors are very similar to the validation ones with average error per parameter around $3.54\%$ for a coarse mesh, $2.89\%$ for a medium mesh and $3.88\%$ for a thin mesh. These results show that the MNN method is capable of estimating the breast model parameters with a very good accuracy in comparison with the error of the IM. For all mesh resolutions the MNN method is clearly more efficient since it provides more accurate approximations in a very short time. The error is reduced by $50\%$ and the computational time reduction, reported in the last column, is very significant.

Clearly, such results highlight the interest to substitute the inverse problem solvers, based on the resolution of the mechanical problem, with a Neural Network at the production level. However, the MNN model estimation can sometimes predict a set of parameters that can notably differ from the exact parameters by a large margin (greater than 30\% error per parameter for instance). On the other hand, the iterative method always provides an accurate estimate within a guaranteed error maximum level (\cite{Lopes2018}) since we really solve the physical problem. Therefore the values that are obtained from the MNN model need to be used with caution. In other words, a validation and correction procedure has to be established to guarantee the validity of the parameters and turn the method more robust.

%%%%%%%%%%%%%%%%%%%%%%%%%%%%%%%%%%%%%%%%%%%%%%
%%%%%%%%%%%%%%%%%%%%%%%%%%%%%%%%%%%%%%%%%%%%%%
\subsubsection{Methods robustness}
\label{subsec:im_vs_nn:robust}

The performance results of the MNN model presented in section \ref{nn:analysis:test} show a good accuracy with relative errors.
However, these tests were done with the MNN models trained with the exact values of the measurements. In a real scenario the surgeon might not obtain the exact measurements of the breast, \textit{i.e.} but an approximation with some inaccuracy and so, it is important to test the robustness of the model, {\it i.e.} the capacity to handle the uncertainties by limiting the error propagation to the outputs.  
From surgeons experience, it is assumed a potential error up to $10\%$ for each measurement. With that in mind, we introduced a variation on each measurement of the test datasets (${\mathcal D}^{test}_{\alpha}$, where $\alpha=coarse, medium, thin$) up to $10\%$.

The sensitivity analysis performed on the estimation method in \cite{Lopes2018} showed that even if the difference on some measurements is small (few percents), it may have a great impact in several situations by strongly magnifying the error on the parameters. Therefore, it is relevant to evaluate how the MNN models deal with potential errors in the measurements.

For that purpose, one introduces errors on the measurements and evaluate the effect of that error in the estimation of the parameters.
To this end, we create a dataset of perturbed measures using a uniform law that produces a maximal error of 10\%. Then we compute the associated output to evaluate the error with respect to the exact value. Hence the relative error of the outputs is evaluated and compared with the relative errors of the inputs. The assessment of the measurements error propagation to the parameters by the MNN is presented in table \ref{table::nn:nn_robust}, together with the error propagation for the IM previously reported in \cite{Lopes2018}. 
\begin{table}[ht]
\setlength{\tabcolsep}{1.5pt}
\renewcommand{\arraystretch}{1.5}
\centering
\begin{tabular}{|c|c|c|c|c|c|c|c|c|}
\hline
\multicolumn{2}{|c|}{\textbf{\begin{tabular}[c]{@{}c@{}}Error per\\ parameter\\ (\%)\end{tabular}}} & \textit{\textbf{$R$}} & \textit{\textbf{$H$}} & \textit{\textbf{$\lambda_{br}$}} & \textit{\textbf{$\lambda_{sk}$}} & \textit{\textbf{$\mu_{br}$}} & \textit{\textbf{$\mu_{sk}$}} & \textit{\textbf{\begin{tabular}[c]{@{}c@{}}avg\\ total\end{tabular}}} \\ \hline
\multirow{2}{*}{\textbf{Coarse}} & \textit{\textbf{IM}} & 8.05 & 9.49 & 15.31 & 19.76 & 6.45 & 39.3 & 16.39 \\ \cline{2-9} 
 & \textit{\textbf{MNN}} & 7.71 & 9.21 & 12.68 & 11.97 & 4.87 & 35.17 & 13.60 \\ \hline
\multirow{2}{*}{\textbf{Medium}} & \textit{\textbf{IM}} & 8.47 & 13.23 & 14.82 & 19.40 & 5.63 & 37.98 & 16.59 \\ \cline{2-9} 
 & \textit{\textbf{MNN}} & 6.68 & 9.09 & 12.29 & 10.58 & 4.67 & 32.36 & 12.61 \\ \hline
\multirow{2}{*}{\textbf{Thin}} & \textit{\textbf{IM}} & 8.96 & 13.41 & 14.70 & 19.17 & 5.21 & 36.08 & 16.25 \\ \cline{2-9} 
 & \textit{\textbf{MNN}} & 7.64 & 11.73 & 12.08 & 11.75 & 4.88 & 32.88 & 13.49 \\ \hline
\end{tabular}
\caption{Mean value of the error in (\%) between the reference parameters and the approximation obtained by the IM and the trained MNN with three different sized meshes considering errors in the measurements up to $10\%$.}
\label{table::nn:nn_robust}
\end{table}

The difference between the estimated parameters after perturbation of the inputs and the exact values increased as expected due to the errors in the measurements. We report a substantial degradation from the original errors $3.51\%$, $2.86\%$ and $3.89\%$ (table \ref{tab:nn:it:rob_50}) to $13.6\%$, $12.61\%$ and $13.33\%$ for the coarse, medium and thin mesh sizes respectively. 

Magnification of the errors is essentially determined by the Jacobian matrix of the functional $M\to \Lambda(M)$ evaluated with the exact measure value $M^{exa}$ ({\it i.e.} without perturbation) and a first-order of errors approximation are given with
$$
\Delta \Lambda_j=\frac{\partial \Lambda_j}{\partial M_i}(M^{exa})\Delta M_i
$$
According to \cite{Lopes2018}, the Jacobian coefficient are uniformly bounded and the error propagation is almost contained leading to reliable values as a first estimator of the breast model parameters. Since the deviations observed with the MNN are very similar to the ones obtained with the IM, we conclude that the MNN is robust.

%%%%%%%%%%%%%%%%%%%%%%%%%%%%%%%%%%%%%%%%%%%%%%%%%%%%%%%%%
%%%%%%%%%%%%%%%%%%%%%%%%%%%%%%%%%%%%%%%%%%%%%%%%%%%%%%%%%
\subsubsection{Measurements deviation}
\label{nn:analysis:applications}

The question arises about the one-to-one correspondence between the measurement and the parameters, namely if the two operations $M\to \Lambda(M)$ (with the IM or MNN) and $\Lambda\to M(\Lambda)$ (with the direct solver) are inverse from one to another at the numerical level. To this end, as in \cite{Lopes2018}, we assess the error between a given reference measurements $\overline{M}$ with the measurements obtained with the estimated parameters $M(\Lambda)$ after computing $\Lambda$ with one of the two method. We evaluate the relative error using
\begin{equation}
\label{eq:nn:cost_measure_m}
E(\overline{M},\Lambda) = \sum_{i=1}^{15} k_{i}\Big ( \overline{M}_{i}-M_i(\Lambda)\Big )^2.
\end{equation}
where $k_{i}$ represents the relative weight of each measurement (the scale of some measurements such as the volume is very different compared to measurements such as the skin surface area) \cite{Lopes2018}. In \cite{Lopes2018}, they considered that a set of parameters is accurately evaluated when the value returned by the cost function in equation \ref{eq:nn:cost_measure_m} is lower than a certain threshold $\varepsilon_{e}=5e^{-6}$ in order to ensure very small errors.

We first start with the IM and evaluate the difference between $\overline{M}$ and $M(\Lambda)$ considering the three mesh sizes. We report in table \ref{table::nn:test_measure_im} the results obtained in \cite{Lopes2018} for all the measurements $M_i$, $i=1,\cdots,15$.

\begin{table}[ht]
\setlength{\tabcolsep}{1.5pt}
\renewcommand{\arraystretch}{1.5}
\centering
\begin{tabular}{|c|c|c|c|c|c|c|c|c|c|}
\hline
\textbf{Mesh} & \textit{\textbf{$e_{m_{1}}$}} & \textit{\textbf{$e_{m_{2}}$}} & \textit{\textbf{$e_{m_{3}}$}} & \textit{\textbf{$e_{m_{4}}$}} & \textit{\textbf{$e_{m_{5}}$}} & \textit{\textbf{$e_{m_{6}}$}} & \textit{\textbf{$e_{m_{7}}$}} & \textit{\textbf{$e_{m_{8}}$}} & \textit{\textbf{$e_{m_{9}}$}} \\ \hline
\textbf{Coarse} & 0.47 & 0.64 & 0.87 & 0.52 & 6.53 & 0.73 & 0.51 & 0.80 & 0.76 \\ \hline
\textbf{Medium} & 0.48 & 0.61 & 0.82 & 0.51 & 6.21 & 0.67 & 0.63 & 0.83 & 0.75 \\ \hline
\textbf{Thin} & 0.48 & 0.50 & 0.79 & 0.51 & 6.18 & 0.61 & 0.51 & 0.85 & 0.73 \\ \hline
\textbf{Mesh} & \textit{\textbf{$e_{m_{10}}$}} & \textit{\textbf{$e_{m_{11}}$}} & \textit{\textbf{$e_{m_{12}}$}} & \textit{\textbf{$e_{m_{13}}$}} & \textit{\textbf{$e_{m_{14}}$}} & \textit{\textbf{$e_{m_{15}}$}} & \textit{\textbf{\begin{tabular}[c]{@{}c@{}}avg\\ total\end{tabular}}} & \multicolumn{2}{c|}{\textit{\textbf{\begin{tabular}[c]{@{}c@{}}cost\\ function\end{tabular}}}} \\ \hline
\textbf{Coarse} & 8.23 & 0.67 & 0.43 & 0.51 & 0.56 & 7.52 & 1.98 & \multicolumn{2}{c|}{1.99e-6} \\ \hline
\textbf{Medium} & 7.11 & 0.73 & 0.52 & 0.57 & 0.55 & 5.87 & 1.79 & \multicolumn{2}{c|}{1.81e-6} \\ \hline
\textbf{Thin} & 7.17 & 0.67 & 0.47 & 0.52 & 0.55 & 5.73 & 1.75 & \multicolumn{2}{c|}{1.72e-6} \\ \hline
\end{tabular}
\caption{Mean value error in \% between the reference and estimated measurements for each mesh using the iterative method (IM). Last lower column shows the mean value of the cost function.}
\label{table::nn:test_measure_im}
\end{table}

We observe an average total error per parameter lower than $2\%$ and we see that, with the exception of $e_{m_{5}}$, $e_{m_{10}}$ and $e_{m_{15}}$, every parameter has an error lower than $1\%$, which indicate an excellent one-to-one correspondence. The error obtained with the other mentioned measurements was also considered small according to \cite{Lopes2018} because their values ranged within the few millimetres. Note that the average error per parameter is slightly better for thinner meshes ($1.98\%$ of error for a coarse mesh and just $1.79\%$ and $1.75\%$ error for the medium and thin mesh respectively).

We performed the same tests using the MNN as a predictor for the parameters reported the errors in table \ref{table::nn:test_measure}.

\begin{table}[ht]
\setlength{\tabcolsep}{1.5pt}
\renewcommand{\arraystretch}{1.5}
\centering
\begin{tabular}{|c|c|c|c|c|c|c|c|c|c|}
\hline
\textbf{Mesh}   & \textit{\textbf{$e_{m_{1}}$}}  & \textit{\textbf{$e_{m_{2}}$}}  & \textit{\textbf{$e_{m_{3}}$}}  & \textit{\textbf{$e_{m_{4}}$}}  & \textit{\textbf{$e_{m_{5}}$}}  & \textit{\textbf{$e_{m_{6}}$}}  & \textit{\textbf{$e_{m_{7}}$}}                                         & \textit{\textbf{$e_{m_{8}}$}}                           & $e_{m_{9}}$                          \\ \hline
\textbf{Coarse} & 0.33                           & 0.33                           & 0.79                           & 0.63                           & 4.63                           & 0.37                           & 0.33                                                                  & 0.76                                                    & 0.60                                 \\ \hline
\textbf{Medium} & 0.22                           & 0.34                           & 0.56                           & 0.48                           & 3.73                           & 0.45                           & 0.52                                                                  & 0.83                                                    & 0.88                                 \\ \hline
\textbf{Thin}   & 0.34                           & 0.36                           & 0.93                           & 0.71                           & 4.43                           & 0.43                           & 0.71                                                                  & 0.87                                                    & 0.92                                 \\ \hline
\textbf{Mesh}   & \textit{\textbf{$e_{m_{10}}$}} & \textit{\textbf{$e_{m_{11}}$}} & \textit{\textbf{$e_{m_{12}}$}} & \textit{\textbf{$e_{m_{13}}$}} & \textit{\textbf{$e_{m_{14}}$}} & \textit{\textbf{$e_{m_{15}}$}} & \textit{\textbf{\begin{tabular}[c]{@{}c@{}}avg\\ total\end{tabular}}} & \multicolumn{2}{c|}{\textit{\textbf{\begin{tabular}[c]{@{}c@{}}cost\\ function\end{tabular}}}} \\ \hline
\textbf{Coarse} & 7.49                           & 0.33                           & 0.28                           & 0.52                           & 0.52                           & 6.08                           & 1.60                                                                  & \multicolumn{2}{c|}{$1.45e^{-6}$}                                                              \\ \hline
\textbf{Medium} & 5.16                           & 0.32                           & 0.44                           & 0.55                           & 0.63                           & 5.31                           & 1.36                                                                  & \multicolumn{2}{c|}{$1.31e^{-6}$}                                                              \\ \hline
\textbf{Thin}   & 7.23                           & 0.29                           & 0.53                           & 0.54                           & 0.66                           & 5.37                           & 1.62                                                                  & \multicolumn{2}{c|}{$1.64e^{-6}$}                                                  \\ \hline
\end{tabular}
\caption{ Mean value error in \% between the reference and estimated measurements for each mesh using the MNN. Last lower column shows the mean value of the cost function.}
\label{table::nn:test_measure}
\end{table}

Results for the MNN (table \ref{table::nn:test_measure}) show smaller errors comparing to the IM (table \ref{table::nn:test_measure_im}).
The average total error per parameter is smaller than $1.62\%$ which is an less amount of error compared to the results obtained with the IM (table \ref{table::nn:test_measure_im}). The error values follow the same pattern presented in tables \ref{table::nn:train_eval_res} and \ref{tab:nn:it:rob_50} with the best results been obtained with the medium mesh. Just like with the IM, apart from $e_{m_{5}}$, $e_{m_{10}}$ and $e_{m_{15}}$ which show higher values of error ($\approx 4\%$, $\approx 7\%$ and $\approx 5.5\%$ respectively), the error obtained with the other measurements is inferior to $1\%$.
 
Concluding, we can state that, when compared to the IM approach, the MNN is capable of achieving a set of parameters which produce a lower error in the retrieved measurements and well-preserved the one-to-one correspondence.

\subsection{Hybrid Method}
\label{sec:hybrid}

Most of the parameters $\Lambda$ provided by the MNN are very good approximations and can be directly used for the breast simulation to determine the shape and behaviour of the breast. However, the MNN model estimation may predict a set of parameters that are not correct, \textit{i.e.} the error in the measurements given by equation \ref{eq:nn:cost_measure_m} was above the accuracy threshold ($5e^{-6}$) which, in practical terms, presents too many errors to produce a relevant breast simulation. This suggests that an assessment of the parameter validity provided by the MNN is critical and correction would be necessary in some very few situations. 

Considering the IM workflow, it uses a pre-determined set of parameters as the initial guess and then proceeds to refine them until the measurements generated from those parameters match the input measurements (generally given by the surgeon - \cite{Lopes2018}). So, as literature shows (\cite{Delingette2004, barabasz2014}), a very good initial guess has a great advantage since the iterative procedure is reduced to very few stages. Despite being reliable, the iterative method can take several minutes (sometimes hours) to estimate the breast parameters. In a medical context, poor estimations can lead to health problems but long running time for parameter estimations are not practical and cannot be used during a consultation. 

\begin{figure}[ht]
\centering
\includegraphics[width=0.30\textwidth]{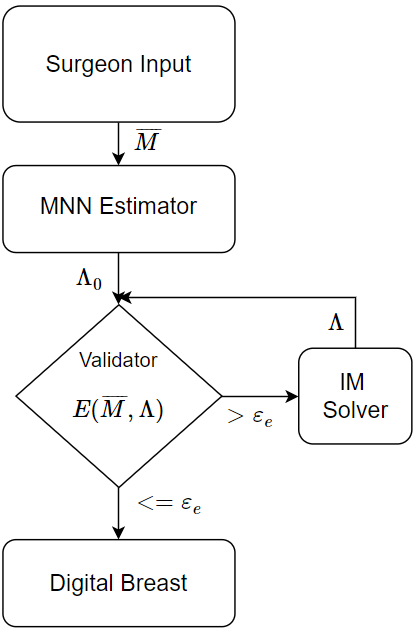}
\caption{Hybrid method.}
\label{fig:hybrid}
\end{figure}

So, to achieve our goal of reliability in an acceptable time frame we propose an hybrid method that takes advantage of the two methods: on the one hand, the quick response of the MNN providing fast and in most cases a good solution for the parameters while, on the other hand, the IM robustness to validate and refine the MNN solution guarantying the overall solution accuracy. The hybrid method starts by running the MNN to obtain an initial estimation. Then it uses the iterative method as a validator (i.e. the cost functional of the inverse problem) to check the error. If the error is above a threshold then the IM uses the initial approximation provided by the MNN as a starting point and iterates until it gets the error below the threshold (see Figure \ref{fig:hybrid}).

If this error is lower to the previously mentioned threshold ($\varepsilon_{e}=5e^{-6}$), then according to \cite{Lopes2018} the estimated parameters are suitable to be used to produce the digital representation of the breast. 

We want to evaluate the performance of this hybrid method on two metrics: 
\begin{itemize}
    \item Fast and accurate solutions - percentage of cases where the MNN obtained an accurate solution by itself;
    \item Validation \& refining performance - reliability of the validation as well as the average number of iterations required by the IM to refine non-accurate MNN estimations.
\end{itemize}

The first metric will reveal at what point can a machine learning method like the MNN replace iterative solvers. The second metric is testing the capacity of the MNN to provide good approximations even when it is not accurate, i.e., we are evaluating the premise behind the observations made in \cite{Lopes2018}, which states that closer initial guesses leads to lesser iterations by the iterative method to obtain an accurate set of parameters.

In summary, we want to determine if under these circumstances, the hybrid method can significantly outperform the IM. In table \ref{table::nn:nn_guess}, we present an error analysis estimated by the MNN model for each mesh size with a total number of $N=10000$ cases per mesh.

\begin{table}[ht]
\setlength{\tabcolsep}{1.5pt}
\renewcommand{\arraystretch}{1.5}
\centering
\begin{tabular}{|c|c|c|c|c|c|}
\hline
\multicolumn{2}{|c|}{\textbf{\begin{tabular}[c]{@{}c@{}}MNN \\ Estimation\end{tabular}}} & \textbf{\begin{tabular}[c]{@{}c@{}}Number\\ of Cases\end{tabular}} & \textbf{\begin{tabular}[c]{@{}c@{}}Results\\ (\%)\end{tabular}} & \textbf{\begin{tabular}[c]{@{}c@{}}Avg\\ Iterations\end{tabular}} & \textbf{\begin{tabular}[c]{@{}c@{}}Time\\ (min)\end{tabular}} \\ \hline
\multirow{4}{*}{\textbf{Coarse}} & \textit{\textbf{Accurate}} & 9887 & 98.9 & - & - \\ \cline{2-6} 
 & \textit{\textbf{Close}} & 82 & 0.8 & 2.3 & 0.62 \\ \cline{2-6} 
 & \textit{\textbf{Medium}} & 18 & 0.2 & 4.5 & 1.22 \\ \cline{2-6} 
 & \textit{\textbf{Far}} & 13 & 0.1 & 7.0 & 1.89 \\ \hline
\multirow{4}{*}{\textbf{Medium}} & \textit{\textbf{Accurate}} & 9907 & 99.1 & - & - \\ \cline{2-6} 
 & \textit{\textbf{Close}} & 71 & 0.7 & 2.2 & 1.96 \\ \cline{2-6} 
 & \textit{\textbf{Medium}} & 15 & 0.2 & 3.6 & 3.20 \\ \cline{2-6} 
 & \textit{\textbf{Far}} & 7 & 0.1 & 7.5 & 6.68 \\ \hline
\multirow{4}{*}{\textbf{Thin}} & \textit{\textbf{Accurate}} & 9873 & 98.7 & - & - \\ \cline{2-6} 
 & \textit{\textbf{Close}} & 77 & 0.8 & 2.4 & 4.54 \\ \cline{2-6} 
 & \textit{\textbf{Medium}} & 30 & 0.3 & 4.8 & 9.07 \\ \cline{2-6} 
 & \textit{\textbf{Far}} & 20 & 0.2 & 7.6 & 14.36 \\ \hline
\end{tabular}
\caption{Neural Network model parameter estimation deviation groups for each mesh size (Coarse, Medium, Thin): Accurate Estimation; Close Estimation; Medium Estimation; Far. Average number of iterations taken from \cite{Lopes2018}, and time in minutes required by the Iterative Method to converge. }
\label{table::nn:nn_guess}
\end{table}

The values presented in table \ref{table::nn:nn_guess} show, for each mesh size, the 4 different types of estimations performed by the MNN trained models: Accurate; Close (difference of parameters up to 10\%); Medium (difference of parameters up to 20\%) and Far (difference of parameters above 20\%). The MNN models are accurate approximately $99\%$ of the times ($98.9\%$, $99.1\%$ and $98.7\%$ with the MNN coarse, medium and thin models respectively). This means that in $99\%$ of the cases, the hybrid model will be able to estimate the breast parameters almost instantaneously. 

From the total of 30000 cases (10000 cases per mesh size) only 333 cases were not accurate: 230 ($0.76\%$ of the cases) were close to the reference values; 63 were medium ($0.21\%$ of the cases) and 40 cases were poor (far) estimations ($0.13\%$ of the cases). 
Note that these cases do not have necessarily extreme measurement values, spreading roughly evenly across the tested parameter spectrum. For these cases we show that the initial guess deriving from the trained MNN, reduces significantly the time spent to obtain an accurate estimation of the parameters. Instead of an average of 3.27, 11.6 and 26.45 minutes for the coarse, medium and thin meshes respectively, the hybrid method results are always significantly below these averages even when the MNN estimation was poor (1.89, 6.68 and 14.36 minutes for the coarse, medium and thin meshes respectively). The hybrid model improve the robustness of the IM method while providing results in a significantly smaller time frame.

\section{Conclusion}
\label{sec:conclusion}

Traditional methods for computing the solution of inverse problems (e.g. estimation of parameters) usually involve iterative solvers and techniques. Such popular techniques require significant computational resources and time response that ranges between a few minutes to several hours, depending on the required accuracy. Machine learning based approaches offer an alternative way to efficiently assess the same problems by providing a solution in real time. However they are not $100\%$ reliable and that limits its usage, mainly when dealing with real problems (e.g. medical) where reliability is crucial.

A new hybrid method was proposed to combine the strengths of the traditional methods (iterative methods) and the new machine learning methods (multilayer neural networks). We used as case study an existent iterative method to estimate the breast parameters of a patient from a set of given measurements. This new method uses data from the problem to create and train a MNN that provides an estimation of the breast parameters. Then, the method uses an IM to validate the estimated parameters and in the case the solution is not considered accurate, then the IM refines those parameters until the desired accuracy is achieved.

The overall results show that the hybrid method proposed achieved its goals: a fast and reliable method. The MNN was capable of providing accurate solution for $99\%$ of cases. On the remaining $1\%$, this method was capable of refining the MNN solution, and when comparing its performance against a stand alone iterative method, we observed that it reduced very significantly the number of iterations required to converge to an accurate solution. 

This new approach of solving inverse problems show the potential of combining MNN together with other numerical methods, as the iterative inverse problem solver, to create more efficient as well as more accurate estimators/predictors for critical scenarios where a non accurate solution is not tolerated. The tests performed with this new model show great potential of its usage in real situations.

\section{Acknowledgements}

A very large thanks to Dra. Augusta Cardoso for her contribution to this paper.

This work has been supported by Portuguese Foundation for Science and Technology (FCT) within the Project Scope: UID/CEC/00319/2019 as well as in the framework of the Strategic Funding UID/FIS/04650/2019.

%%%%%%%%%%%%%%%%%%%%%%%%%%%%%%%%%%%%%%%%%%%%%
%%%%%%%%%%%%%%%%%%%%%%%%%%%%%%%%%%%%%%%%%%%%%
%%%%%%%%%%%%%%%%%%%%%%%%%%%%%%%%%%%%%%%%%%%%%
%%%%%%%%%%%%%%%%%%%%%%%%%%%%%%%%%%%%%%%%%%%%%


\bibliographystyle{elsarticle-num}
\bibliography{refs}






% \end{thebibliography}

\end{document}